\documentclass[aps,prb,twocolumn,showpacs,10pt]{revtex4-1}
\usepackage[dvips]{graphicx}
\usepackage{mathrsfs}
\usepackage{amsmath}
\usepackage{amssymb}
\usepackage{amsfonts}
\usepackage{ae}
\usepackage{bm}
\usepackage{units}
\usepackage{natbib}

\usepackage{xcolor}
\usepackage{hyperref}
\hypersetup{
    colorlinks,
    linkcolor={blue!75!black!80!yellow},
    citecolor={blue!75!black!80!yellow},
    urlcolor={blue!75!black!80!yellow}
}

\begin{document}
%

\title{Localized plasmons in bilayer graphene nanodisks}
\author{Weihua Wang}
\thanks{Present Address: Department of Physics, China University of Mining and Technology, Xuzhou 221116, China}
\affiliation{DTU Fotonik, Department of Photonics Engineering, Technical University of Denmark, DK-2800 Kongens Lyngby, Denmark}
\affiliation{Center of Nanostructured Graphene (CNG), Technical University of Denmark, DK-2800 Kongens Lyngby, Denmark}
\author{Sanshui Xiao}
\thanks{Email: saxi@fotonik.dtu.dk}
\affiliation{DTU Fotonik, Department of Photonics Engineering, Technical University of Denmark, DK-2800 Kongens Lyngby, Denmark}
\affiliation{Center of Nanostructured Graphene (CNG), Technical University of Denmark, DK-2800 Kongens Lyngby, Denmark}
\author{N. Asger Mortensen}
\thanks{Email: asger@mailaps.org}
\affiliation{DTU Fotonik, Department of Photonics Engineering, Technical University of Denmark, DK-2800 Kongens Lyngby, Denmark}
\affiliation{Center of Nanostructured Graphene (CNG), Technical University of Denmark, DK-2800 Kongens Lyngby, Denmark}
\date{\today}
%
\begin{abstract}
We study localized plasmonic excitations in bilayer graphene (BLG) nanodisks, comparing AA-stacked and AB-stacked BLG and contrasting the results to the case of two monolayers without electronic hybridization. 
The electrodynamic response of the BLG electron gas is described in terms of a spatially homogeneous surface conductivity, and an efficient alternative two-
dimensional electrostatic approach is employed to carry out all the numerical calculations of plasmon resonances. Due to a unique electronic band structures, the resonance frequency of the traditional dipolar plasmonic mode in the AA-stacked BLG nanodisk is roughly doping independent in the low-doping regime, while the mode is highly damped as the Fermi level approaches the interlayer hopping energy $\gamma$ associated with tunneling of electrons between the two layers. In addition to the traditional dipolar mode, we find that the AB-stacked BLG nanodisk also hosts a new plasmonic mode with energy larger than $\gamma$. This mode can be tuned by either the doping level or structural size, and furthermore, this mode can dominate the plasmonic response for realistic structural conditions.
\end{abstract}
\pacs{73.20.Mf, 71.35.Ji, 78.67.Wj}

\maketitle

\section{Introduction}
Graphene, a flat two-dimensional (2D) crystal, is made of a single layer of carbon atoms arranged in a honeycomb lattice with planar $sp^2$ hybridized orbitals. The remaining out-of-plane $p_z$-orbitals form delocalized $\pi$ bands. Unlike two-dimensional electron gases (2DEGs) in traditional semiconductors, the quasi particles in graphene $\pi$ bands act as gapless Dirac fermions with high carrier mobility,\cite{rmp_81_109, rmp_83_407} leading to a number of unique optical properties such as a universal 2.3\% linear light absorption in the visible regime.\cite{science_320_1308} In the infrared regime, the optical properties of doped graphene are largely determined by the collective excitations of $\pi$ electrons (plasmons), which exhibit distinct features in doped graphene, including long propagation length, extreme mode confinement and field enhancement. In particular, plasmon resonance frequencies are tunable through chemical doping or electrostatic gating.\cite{nl_11_3370, nature_487_77, nature_487_82, acsphotonics_1_135, front_phys_11_117801} The so-called graphene plasmons have already been demonstrated to facilitate quite diverse light-matter interaction phenomena, such as modified emitter-radiation dynamics,\cite{prb_84_195446} wave propagation,\cite{acsnano_6_431} energy absorption,\cite{prb_85_081405, prl_108_047401} and possibly also plasmomechanics.\cite{nanoscale_8,natcommun_7_10218} This opens up a new avenue towards future plasmonic applications,\cite{acsnano_8_1086} and will stimulate a broad exploration of plasmonics in 2D systems,\cite{jpcm_26_123201} in which bilayer graphene is one exciting example.\cite{prb_82_195428, prb_85_075410, prb_88_115420, njp_14_10-105018, prl_112_116801, nanophotonics_4_115}

\begin{figure}[b]
\includegraphics[width=8.0cm]{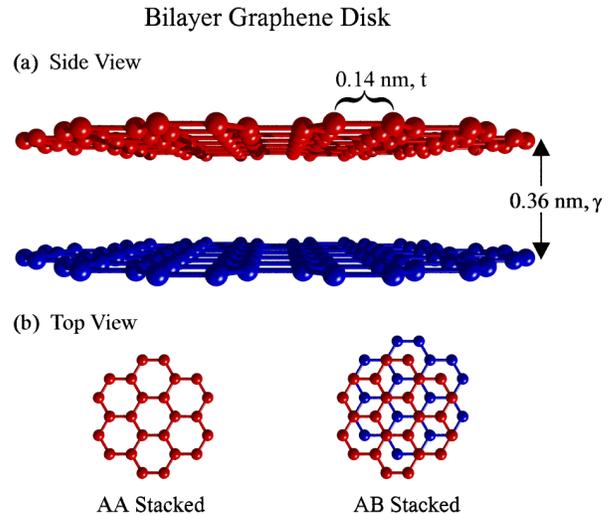}
\caption{(color online) Schematic diagrams for bilayer graphene disks, showing both a side view (a) and a top view (b) with two typical stacking sequences; exactly aligned (AA stacked) and relatively translated (AB stacked).}
\end{figure}

Bilayer graphene (BLG) is composed of two layers of graphene stacked under certain sequences such as AA-stacking and AB-stacking, see Fig.~1. The AA-stacked BLGs are exactly aligned, and while some theory models predict them to be structurally unstable due to shear layer shift,\cite{prl_109_206801} they have recently nevertheless been produced from thermally treated graphite.\cite{prl_102_015501} Because of the predicted instability, the AA-stacked BLGs have been less intensively studied so far. However, the AA-stacked BLGs are interesting in their own right, for instance being metallic even at zero doping with two conducting bands crossing at zero energy.\cite{prb_88_115420, prl_109_206801} Differing from the AA-stacked BLGs, the AB-stacked BLGs can be thought as the twisted bilayers with respect to AA-stacked BLGs by rotating 60 degrees with respect to the axis of sublattice AA in each unit cell, \cite{prl_99_256802} or very straightforwardly, by shifting one of the layers by a vector $(\sqrt{3}, 1)a/2$ where $a\approx 0.14$\,nm is the carbon-carbon bond length. The AB-stacked BLGs are most commonly studied \cite{science_313_951, rpp_76_056503} and the have potential applications in electronic and opto-electronic devices owing to the tunability of the band gap.\cite{nature_459_820,  nanophotonics_4_115}

From the tight-binding point of view, the unique electronic properties of BLGs are determined by the intralayer hopping energy $t$ and the interlayer hopping energy $\gamma$ of $p_z$-orbitals.\cite{prb_80_165406} Since the interlayer distance $d\approx 3.6$ nm is $2.5$ times larger than the carbon-carbon bond length $a$, $\gamma$ is naturally much smaller than $t$. As such, the interlayer coupling is mainly perturbing the Dirac dispersion properties associated with the monolayer graphene (characterized by a fixed $t$), while the interlayer coupling $\gamma$ varies slightly for the different bilayers. Usually $\gamma$ is a little smaller in AA-stacked BLGs than in AB-stacked BLGs, which is also one of the reasons why AB-stacked BLGs are more stable. In the continuum limit we consider the momentum in the vicinity of the $K$ point of the Brillouin zone (where the Dirac equation captures the dynamics of the $\pi$ electrons\cite{rmp_81_109}), the interlayer coupling leads to four low energy electronic bands. As for monolayer graphene, also here the energy levels and associated wave functions form the starting point for calculations of optical excitations based on linear response theory,\cite{njp_8_318, prb_75_205418} from which the macroscopic surface conductivity is derived -- a quantify that forms a starting point for subsequent explorations of optical properties.\cite{prb_88_115420, prb_77_155409} In this way, plasmons in monolayer graphene have been widely explored.\cite{acsphotonics_1_135,front_phys_11_117801} However, plasmon phenomena in BLGs nanostructures are still to be investigated. Clearly, ideal infinite BLG sheets are already interesting,\cite{prb_88_115420, prl_112_116801, oe_19_11236} while finite and artificially structured flakes constitute an unexplored territory.

\begin{figure}[t!]
\includegraphics[width=\columnwidth]{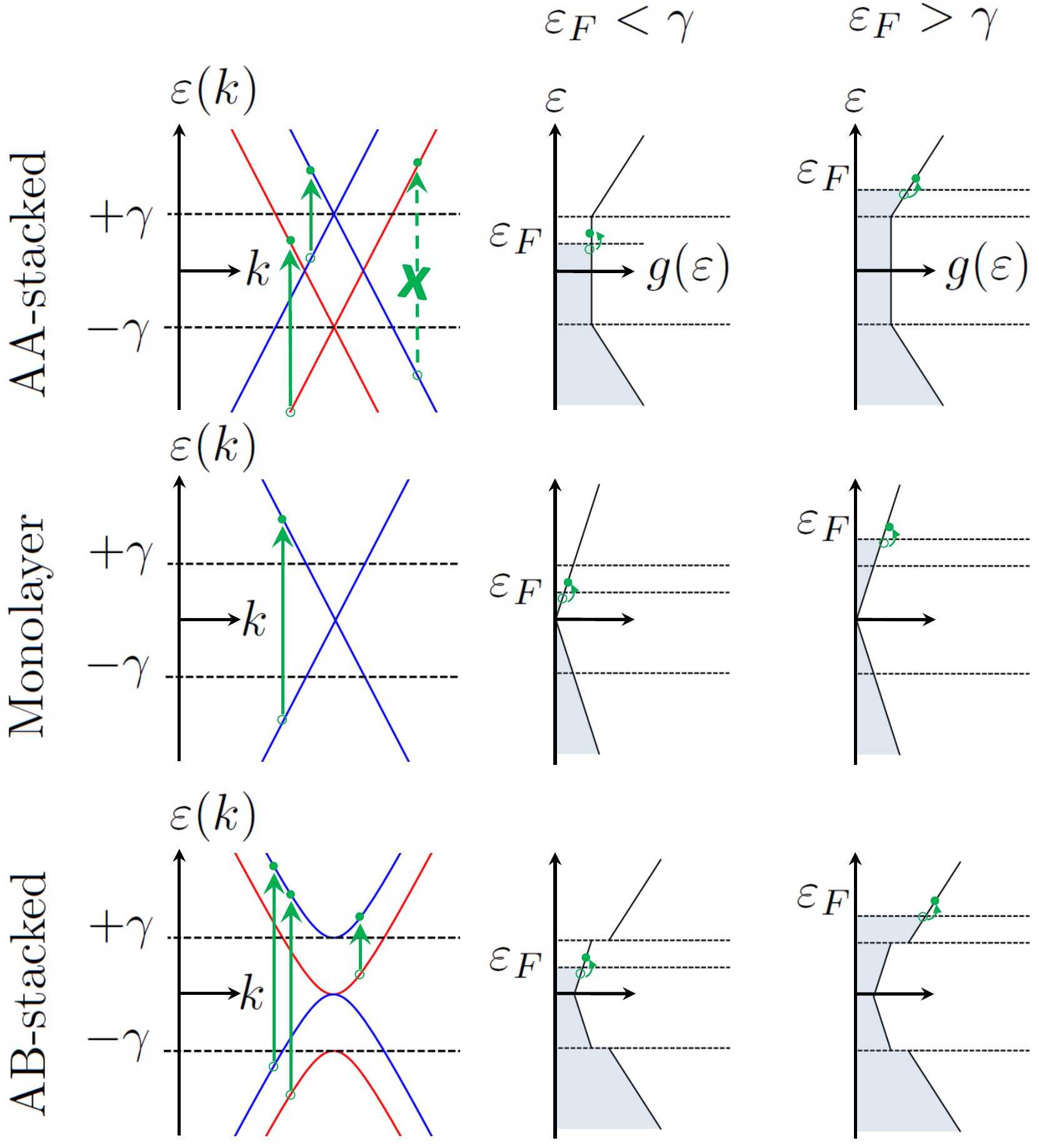}
\caption{(color online) Schematic electronic band structures $\varepsilon(k)$ [left column] and electronic density-of-states $g(\varepsilon)$ [middle and right columns] of AA-stacked, monolayer graphene, and AB-stacked bilayer graphene. The arrows indicate possible and prohibited (dashed style arrow) single-particle interband transitions (vertical transitions) as well as intraband plasmonic excitations near the Fermi level $\varepsilon_F$.}
\end{figure}

In this paper we investigate the localized plasmon excitations in the AA-stacked BLG, the AB-stacked BLG, and also double monolayer graphene (MLG) nanodisks as a comparison. We adopt here the double MLG being similar to BLG shown in Fig.~1(a), but without tunneling between layers. The light-matter interactions are treated at the macroscopic level, where graphene is characterized by a homogeneous surface conductivity. As is common practice,\cite{acsphotonics_1_135,front_phys_11_117801} we employ the conductivities of infinite graphene sheets derived from the framework of the random-phase approximation (RPA), \cite{njp_8_318, prb_75_205418, jap_103_064302} where the quantum nature of electrons and nonlocal effects in finite systems are ignored as we focus on the structures with feature size of tens of nanometers.\cite{ncom_5_3809, ncom_5_3548, acsnano_6_1766, prb_90_241414} Likewise, substrate phonons and nonlinear effects are ignored as well. \cite{nl_14_2907, prb_92_121407} In practice, the surface conductivities are taken to be frequency-dependent functions, including the contributions from both intraband and interband transitions. An efficient numerical approach employing a two-dimensional finite-element method was developed in our previous work, \cite{scirep_5_9535} and here it is employed with bilayer conductivity expressions to perform all the calculations. We report both optical absorption spectra of the three structures and the extracted band diagrams, which illustrate the different plasmonic behaviors as a function of Fermi levels. As a particular novelty, we find that while plasmonic resonance frequencies usually exhibit a simple square-root dependence, the dispersion is more complicate in both AA-stacked and AB-stacked BLG nanodisks. The complexity originates from the existence of two further branches in the electronic band structure, and thus more transition processes contribute to the surface conductivities (see Fig.~2). We find that the AA-stacked BLG nanodisks support localized plasmon excitations even at zero doping. The plasmonic frequency decreases when the doping level increases up to $\gamma$, and the plasmons will be damped out when exceeding $\gamma$. In the AB-stacked BLG nanodisks, a new plasmon mode at higher frequency will arise as the doping level exceeds $\gamma$. Quite interestingly, this mode will dominate when the Fermi level exceeds $2\gamma$. These interesting phenomena have no counterparts in monolayers and stem from the additional single-particle intraband and interband transitions involving the two new branches in the electronic band structure. Especially in the AA-stacked BLGs, interband transitions turn out to play an important role.

\section{Surface conductivities of BLGs}
The optical response of graphene is dominated by the single particle transitions among low energy bands ($\pi$ and $\pi^*$ bands), as indicated by the vertical arrows in Fig.~2. In the AA-stacked BLGs there are two Dirac cones crossing at $\epsilon=0$, each being shifted up/down in energy by $\gamma$. As a consequence of the symmetry requirement, the single particle transitions between different Dirac cones in the AA-stacked BLGs are prohibited,\cite{prb_88_115420} and thus the surface conductivity is composed from the two independent cones. Thus, surface conductivities of the AA-stacked BLGs remain relatively simple. For instance, if the surface conductivities of MLGs are $\sigma(\omega, |\epsilon_F|)$ as demonstrated recently,\cite{njp_8_318, prb_75_205418, prb_80_245435} then the surface conductivities of the AA-stacked BLGs will be proportional to $\sigma(\omega, |\epsilon_F+\gamma|)+\sigma(\omega, |\epsilon_F-\gamma|)$. As an example, for AA-stacked bilayers the complex-valued conductivity is\cite{jetpl_97_429} 
\begin{equation}
\begin{split}
\sigma_{\rm AA}(\tilde{\omega})&=\frac{ie^2}{2\pi\hbar}\left[\frac{|\epsilon_F+\gamma|}{\hbar\tilde{\omega}}+\frac14\ln\frac{2|\epsilon_F+\gamma|-\hbar\tilde{\omega}}{2|\epsilon_F+\gamma|+\hbar\tilde{\omega}}\right]\\
&+\frac{ie^2}{2\pi\hbar}\left[\frac{|\epsilon_F-\gamma|}{\hbar\tilde{\omega}}+\frac14\ln\frac{2|\epsilon_F-\gamma|-\hbar\tilde{\omega}}{2|\epsilon_F-\gamma|+\hbar\tilde{\omega}}\right],
\end{split}
\end{equation}
where $\tilde{\omega}=\omega+i\tau^{-1}$ is the complex frequency including and imaginary part associated with a phenomenological relaxation time $\tau$. The first term in the two brackets is the intraband Drude term, while the second term accounts for the interband transitions. In passing, we note a quite interesting observation for the Drude model with $\sigma(\omega, |\epsilon_F|)$: if $|\epsilon_F|<\gamma$ then the surface conductivity of the AA-stacked case in Fig.~2 does not depend on the particular $\epsilon_F$. Including interband transitions as discussed below, we return to a doping-dependent expression and as a result both the resonance frequency and the resonance linewidth depend on doping.

The surface conductivity of AB-stacked BLGs can be derived by using a similar approach as employed for MLGs, but additional single particle transitions should be included in the calculations. Fig.~2 illustrates additional allowed transitions between the different band branches. We notice that there are two branches touching at the charge-neutrality point, so that there are no free carriers that can contribute to intraband plasmon excitations. Furthermore, the dispersion relations deviate from the linear relationship commonly associated with graphene. This is due to the small, yet finite probability for interlayer tunneling of electrons. The band gap between the two positive energy branches is $\gamma$. Thus if $|\epsilon_F|>\gamma$, the upper branch will offer new intraband transitions (see lower, rightmost panel in Fig.~2), leading to a new intraband plasmon mode at higher energy. The surface conductivity of the AB-stacked BLGs should include these new intraband transitions, and fortunately their analytical expressions have already been obtained.\cite{prb_77_155409, prb_80_241402, prb_82_195428} Here, we reproduce the expressions given by Eqs.~(4)--(6) in Ref.~\onlinecite{oe_19_11236},

\begin{widetext}
\begin{multline}
\text{Im}[\sigma_{\rm AB}(\omega)]=\frac{e^2}{2\pi\hbar}\times\Big\{
f(\hbar\omega,2\epsilon_F)+g(\hbar\omega,\epsilon_F,\gamma)+h_+(\hbar\omega,\epsilon_F,\gamma)
\\
+\Theta(\gamma-\epsilon_F)
\left[
f(\hbar\omega,2\gamma)+f(\hbar\omega,\gamma)\times(\gamma/\hbar\omega)^{2} +g(\hbar\omega,\gamma,-\gamma)+(\gamma/\hbar\omega)
\right] \\
+\Theta(\epsilon_F-\gamma)
\left[
f(\hbar\omega,2\epsilon_F)+f(\hbar\omega,2\epsilon_F-\gamma) +g(\hbar\omega,\epsilon_F,-\gamma)+h_{-}(\hbar\omega,\epsilon_F,\gamma)
\right]
\Big\}
,
\end{multline}
\end{widetext}
where $\Theta(\ldots)$ is the Heaviside function, while the other dimensionless functions are given by
\begin{equation}
\nonumber
\begin{split}
&f(x,y)=\frac12\ln\left|\frac{x-y}{x+y}\right|, \\
&g(x,y,z)=\frac{z}{2(x^2-z^2)}\left[x\ln\frac{|x^2-4y^2|}{|2y+z|^2}+z\ln\left|\frac{x+2y}{x-2y}\right| \right], \\
&h_{\pm}(x,y,z)=\frac{2y\pm z}{x}+\frac{xz}{x^2-z^2}\ln\frac{2y\pm z}{z}. \\
\end{split}
\end{equation}
Although this expression appears complex, numerical implementations are straight forward, and the corresponding real part of the conductivity $\text{Re}[\sigma_{\rm AB}(\omega)]$ can be obtained through the Kramers--Kronig relation.

\section{Numerical method}
Using the surface conductivities of BLGs, we now calculate the localized plasmon resonances associated with finite BLG flakes. In MLG structures, it has been common practice to model the 2D graphene sheet as a very thin three-dimensional (3D) film with the artificial thickness $t_g$ associated with the effective bulk permittivity through $\varepsilon(\omega)=\varepsilon_0+i\sigma(\omega)/\omega t_g$.\cite{science_332_1291} While this approach is intuitive for implementation in existing 3D electrodynamics solvers, it requires tremendous computational resources in terms of memory and time. However, such an approach would numerically become even more challenging in BLGs because the artificial thickness $t_g$ should be chosen much smaller than the layer separation of 0.36\,nm. Instead, we follow our previous work, \cite{scirep_5_9535} where all evaluations are restricted to the conducting plane (assumed infinitely thin), where the potential $\phi(\bm r)$ and the charge density $\rho(\bm r)$ are related by
\begin{subequations}
\begin{align}
&\phi(\bm r)=\phi_{\rm ext}(\bm r)+\frac{1}{4\pi\varepsilon_s}\int_{2D}d\bm r^{\prime}\frac{\rho(\bm r^\prime)}{|\bm r-\bm r^\prime|}, \\
&\rho(\bm r)= \frac{i\sigma(\omega)}{\omega}\nabla_{2D}^2\phi(\bm r).
\end{align}
\end{subequations}

\begin{figure*}[t]
\includegraphics[width=18.0cm]{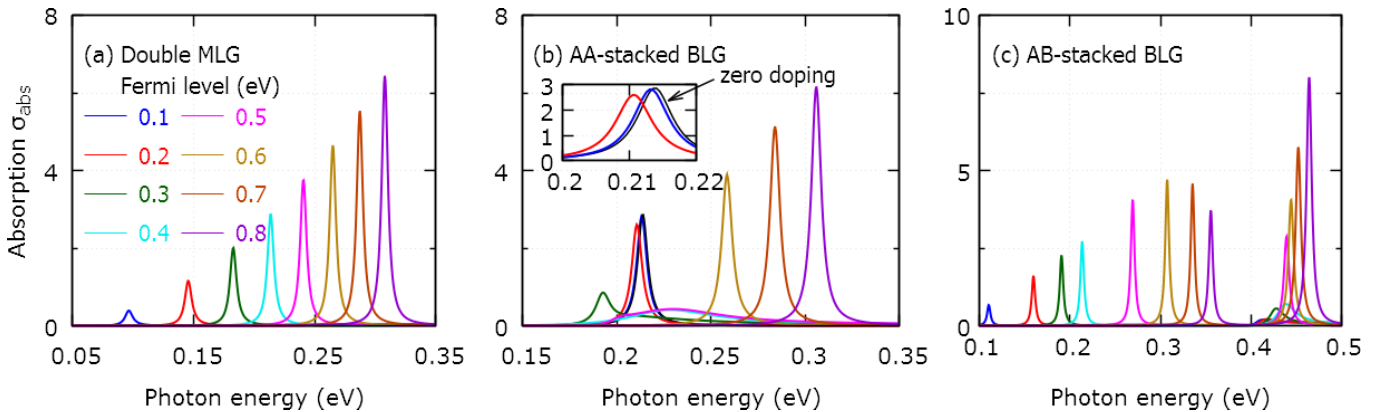}
\caption{(color online) Absorption spectra for varying Fermi levels for (a) the double MLG nanodisks, (b) the AA-stacked BLG nanodisks, and (c) the AB-stacked BLG nanodisks. The inset in (b) shows the magnification of absorption for zero doping (black line), and $\epsilon_F=0.1$\,eV  and 0.2\,eV. The disk radius is $R=50$\,nm and the relaxation loss is $\hbar\tau^{-1}=6$\,meV.}
\end{figure*}

Equations~(3a) and (3b) are self-consistent equations, where $\int_{2D}$ and $\nabla_{2D}^2$ are two-dimensional integral and Laplace operators respectively, $\varepsilon_s$ denotes the average dielectric constant of the surrounding medium, and $\phi_{\rm ext}(\bm r)$ is the external potential. Setting $\phi_{\rm ext}(\bm r)=0$, Eqs.~(3a) and (3b) can be cast into an eigenvalue problem, and the eigenvalues and eigenvectors obtained represent the frequencies and induced charge densities of the plasmonic eigenmodes.\cite{scirep_5_9535} There are some obvious advantages of such a method. First, all modes are obtain at once. Furthermore, the calculation provides us with both bright and dark modes. In this paper, we focus on dipolar plasmonic modes, by investigating the absorption spectra under a plane-wave illumination. The BLG nanodisks lie in the $xy$-plane, and the plane wave $E_{\rm ext}e^{-i\omega t}\hat{x}$ is incident normally. Based on Eqs.~(3a) and (3b), one can calculate the induced charge density $\rho(\bm r)$ for the external excitations $\phi_{\rm ext}(\bm r)=-xE_{\rm ext}$ (please refer to Ref.~\onlinecite{scirep_5_9535} for more details of the numerical procedure), and then the dipole polarizabilities $\alpha(\omega)=\int_{2D}x\rho(\bm r)d\bm r/E_{\rm ext}$. The normalized absorption coefficient is given by
\begin{equation}
\sigma_{\rm abs}=\frac{k_0}{\varepsilon_0S}\text{Im}\{\alpha(\omega)\},
\end{equation}
which is the absorption cross section normalized by the surface area $S$. Here, $k_0=\omega/c$ the wave vector in vacuum. For simplicity, but without loss of generality, we choose $\varepsilon_s = \varepsilon_0$ (structures embedded in vacuum) and for the relaxation loss we use $\hbar\tau^{-1}=6$\,meV throughout this work.\cite{} In order to enable faster convergence of the calculations, the triangular meshes in both layers have been made to be exactly identical. Technically, we create triangular meshes in one layer and then copy all the meshes to the second layer. Due to the symmetry, a high accuracy can be achieved using a relatively lower mesh density, for example 3000 triangles in each layer for disks of radius $R=50$\,nm.

\section{Results and Discussions}

In Fig.~3, we show the absorption coefficient $\sigma_{\rm abs}$ for nanodisks with stacking of the three kinds considered in Fig.~2. In the double MLG nanodisks, the plasmonic frequency increases gradually as we  increase the Fermi levels. Because of very tiny separation between layers, there are very strong electromagnetic interactions despite no electronic coupling. A clear evidence is the plasmonic frequency shift relative to MLG nanodisks. Taking a Fermi level $\epsilon_F=0.5$\,eV and $R= 50$\,nm as an example, the dipolar plasmonic mode $\omega_p=0.173$\,eV in MLG nanodisks and $\omega_p=0.241$\,eV in double MLG nanodisks. While we focus our attention on bright plasmonic modes, we note that for dark plasmonic modes (dipoles in the two layers aligned anti-parallel) can be easily be studied using the same plasmonic eigenmodes approach.\cite{scirep_5_9535} As has pointed out previously,\cite{prb_87_195424} the interband transitions will decrease the plasmonic frequency $\omega_p$, and the contribution from interband transitions will be larger for a larger $\omega_p$ at a given Fermi level. Thus, in double MLGs the deviation from the square root is more apparent.

Another very interesting phenomenon is the variation of plasmonic frequency in the AA-stacked BLG nanodisks. It can be seen in Fig.~3(b) that there is a well-defined plasmonic peak even when $\epsilon_F=0$ and that it almost coincides with the plasmonic peak when the doping is changed to $\epsilon_F=0.1$\,eV, see the blue line. This is a unique property of the AA-stacked BLGs, while both double MLGs and AB-stacked BLGs do not host plasmonic resonances at zero doping. Moreover, while the doping level is relatively small, for instance much smaller than $\gamma=0.4$\,eV, the plasmonic frequency hardly changes with doping. At small doping levels, the intraband transitions dominate and as a result the surface conductivity exhibits a Drude-like behavior. In numerical calculations, we show this behavior persists until $\epsilon_F\simeq 0.1$\,eV. As discussed above, the larger plasmonic frequency will also lead to larger interband contribution, so it is clear that this behavior will eventually break down as we turn to smaller nanodisks. As increasing the doping levels beyond $0.1$\,eV, the plasmonic mode will exhibit a distinct frequency redshift along with a broadening of the peak. In Fig.~2, it is indicated that due to different band filling, the two Dirac cones will not contribute equally to the interband transitions, and when the Fermi level approaches the Dirac point of the upper cone, the upper cone plays a major role. Thereby, both the frequency shift and peak broadening are determined by the upper cone. Especially when $\epsilon_F=\gamma$, the upper cone will only have interband transitions, which leads to very large damping in the plasmonic resonances. However based on this point of view, it is difficult to understand the different damping at $\epsilon_F=0.3$\,eV and $\epsilon_F=0.5$\,eV since they are completely symmetrical relative to the Dirac point. Again, this originates from the larger plasmonic frequency at $\epsilon_F=0.5$\,eV. By increasing the doping level further, the plasmonic frequency increases gradually which is similar to that observed in the double MLG nanodisks.

In the AB-stacked BLG nanodisks, there are no free carriers to support a plasmonic resonance in the absence of doping. With a finite doping, a pronounced plasmonic peak shows up and similar to the case for the double MLG nanodisks, its frequency increases with the increasing of the Fermi level. However, different from the double MLG and the AA-stacked BLG nanodisks, there appears an extra plasmonic mode at higher energy, typically when its frequency $\omega_p>\gamma$. Turning to Fig.~2, the upper-branch plasmonic mode in AB-stacked structures (lower, rightmost panel) arises from the intraband transitions involving the highest branch in the electronic band structure. In Fig.~3(c), we notice that at a low Fermi level ($\epsilon_F<\gamma=0.4$\,eV), the traditional dipole resonance dominates in the absorption spectra. However, the intensity of upper-branch mode increases dramatically at higher Fermi levels ($\epsilon_F>\gamma$). In our calculations shown in Fig.~3(c), the onset of the upper-branch plasmon mode is clearly seen in the $\epsilon_F=0.5$\,eV curve (shown in magenta), where the resonance appears at a photon energy of $\hbar\omega\simeq 270$\,meV. The strength of the upper-branch plasmon mode is comparable to the intensity of the traditional one when $\epsilon_F>\gamma$, and the strength can even be larger when increasing the doping to $\epsilon_F=0.7$\,eV and $0.8$\,eV. In fact, the frequency can be elevated further, for example by applying a bias voltage to open a gap at $\epsilon=0$. In addition to modulating the Fermi levels and band structures, there is also the possibility for geometrical size tuning of the resonance. While it is a commonly used method to shift spectra, we note that is can also alter the superposition of eigenstates to the excited. We will discuss this in more detail below. The peculiar property of the upper-branch mode could be used to extend the plasmonic frequency region in graphene-based plasmonic devices.

\begin{figure}[b]
\hspace*{-0.8cm}\includegraphics[width=9.9cm]{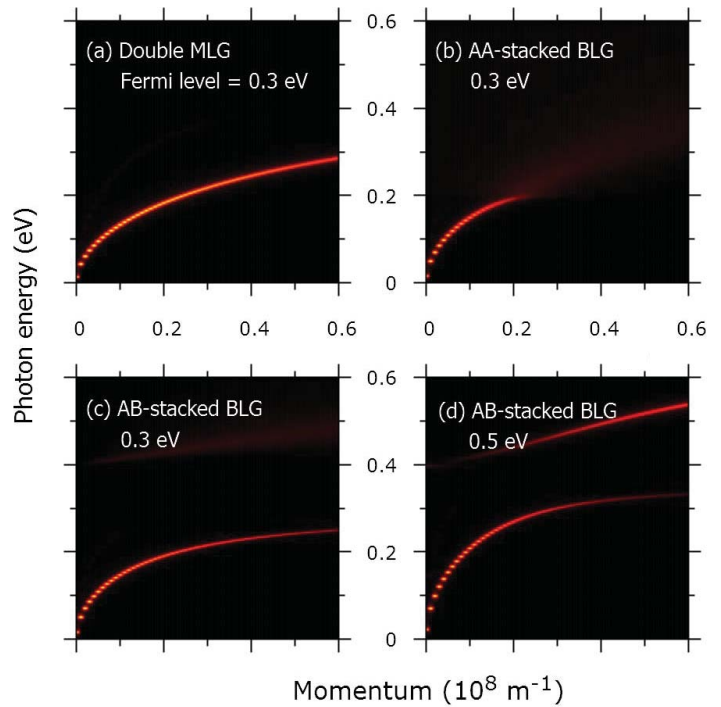}
\caption{(color online) The extracted plasmonic dispersion band diagrams at $\epsilon_F=0.3$\,eV of (a) the double MLG nanodisks, (b) the AA-stacked BLG nanodisks, (c) the AB-stacked nanodisks (c); and (d) the AB-stacked nanodisks for $\epsilon_F=0.5$\,eV.}
\end{figure}
As discussed above, plasmonic frequencies in graphene nanostructures depend on the structural size, and this holds even within the electrostatic approximation, which is different from the physics of dipole resonances in three-dimensional metal particles.\cite{acsnano_8_1745} In metals, the electrostatic approximation will lead to constant plasmonic frequency for the same geometrical structure, but with different sizes. However in graphene, plasmonic frequencies show a typical dependence on the size, $\omega_p\propto\sqrt{R^{-1}}$, which can be easily explored in nanodisks.\cite{prb_86_125450} In two-dimensional disks, the whispering-gallery modes traveling along the perimeter follow a simple relationship
\begin{equation}
n\lambda_p = 2\pi R,\quad n=1, 2, 3, \ldots
\end{equation}
where $n$ is the number of standing waves and $\lambda_p$ is the plasmonic wavelength. From this equation, we see that the wave vector is  $k_p=2\pi\lambda_p^{-1}=nR^{-1}$, where $n=1$ for the dipolar plasmonic mode. Thus, one can reconstruct the dispersion band diagram using e.g. resonance energies for the dipolar plasmonic mode in differently sized disks (for the use in analyzing experiments, see e.g. Ref.~\onlinecite{nl_14_2907}). In practice, we calculate a set of the absorption coefficients for the nanodisks with varying radius, and then organize all the data to create an absorption-intensity color map.

Figure~4 illustrates the plasmonic dispersion band diagrams. One can see that the plasmonic mode in the double MLG nanodisks is always well defined in the calculated regime, but as increasing the wave vector (corresponding to reduce the radius) the intensity decreases slightly. However, the behavior is qualitatively different in the AA-stacked BLG nanodisks. As the wave vector gets larger than $0.2\times10^{8}$\,m$^{-1}$ (equivalent to the radius being smaller than 50\,nm), the plasmonic mode is nearly damped out. The difference here can be understood in the following way. The interband transitions result in plasmonic damping, and the quantity depends on the ratio between the plasmonic frequency $\omega_p$ and twice the Fermi energy $2\epsilon_F$. In the double MLG nanodisks, the ratio increases as increasing the wave vector and is roughly up to $2/3$ at $k_p=0.6 \times 10^{8}$\,m$^{-1}$. In the AA-stacked BLG nanodisks when $\epsilon_F=0.3$\,eV, the upper Dirac cone (see Fig.~2) will have an effective Fermi level $|\gamma-\epsilon_F|=0.1$\,eV, and thereby the ratio could be larger than 1, where the damping from interband transitions is strong enough to disrupt the coherence of the collective excitations.
This is the mechanism that the plasmons fade out when $k_p>0.2\times 10^{-8}$\,m$^{-1}$. The damping mechanism in the AB-stacked BLG nanodisks is quite different. When $\epsilon_F<\gamma$, it is easy to find that (see Fig.~2) there is a new damping path from the interband transition between the two upper band branches. The strength of this damping is determined by a new ratio $\omega_p/\gamma$, and it dominates when $\gamma<2\epsilon_F$. This can be used to explain the larger loss in the AB-stacked BLG nanodisks than in the double MLG nanodisks at $\epsilon_F=0.3$\,eV. Apart from the damping route, there is a new plasmonic dispersion band above the energy $\gamma$, see Fig.~4(c) but it is too weak. To make use of this new mode, one needs to enhance the strength of its resonance. As discussed above, increasing the doping level is a possible way, where this mode will be dominate when $\epsilon_F\ge 0.8$\,eV. This doping level probably is too high to be reached. As we know, the intraband transitions are required for the collective plasmonic excitations. Thus it is very natural to aim for a Fermi energy $\epsilon_F>\gamma$, where the upper branch in the bandstructure is populated. As shown in Fig.~4(d) at an experimentally accessible Fermi energy $\epsilon_F=0.5$\,eV, the new upper-branch plasmonic mode can be enhanced by reducing the radius of the nanodisks. In reality, one should balance the two aspects to optimize the performance.

To conclude, we have studied the plasmonic properties in the AA-stacked BLG nanodisks and the AB-stacked BLG nanodisks. We have found that their plasmonic excitations show qualitatively different behavior. In the AA-stacked BLG nanodisks, there is a well defined plasmonic resonance even without doping, and the plasmonic frequency does not change at small doping level. However, as the doping approaches the energy $\gamma$, there is a very strong damping through interband transitions. This effect would be much stronger at smaller nanodisks where the plasmonic frequency is quite large. In the AB-stacked BLG nanodisks, a new plasmonic mode will emerge at an energy larger than $\gamma$, and the strength of the mode can be tuned by either the doping level and the structural size. Comparing to the traditional plasmonic mode, this mode can be dominate the absorption spectrum when increasing the doping level and shrinking the structural size.
Here, we focused on single disks, but graphene dimers\cite{lpr_7_297,prb_92_205405,scirep_5_9535} naturally constitute another interesting direction for bilayer graphene plasmonics.

\section{Acknowledgements}
We thank Wei Yan, Martin Wubs, and Antti-Pekka Jauho for stimulating discussions. The Center for Nanostructured Graphene (CNG) is sponsored by the Danish National Research Foundation, Project DNRF103. We also acknowledge the Danish Council for Independent Research (FNU 1323-00087).

\bibliographystyle{apsrev4-1}
\bibliography{Database}

\end{document}